\documentstyle[preprint,aps]{revtex}
                                     
\begin{document}                                                                
                                                                                
\def \vsp {\vspace{.3cm}}
\def \vss {\vspace{.4cm}}                                                       
\def \vsl {\vspace{1cm}}                                                        
\def \fl {\flushleft}                                                           
\def \fr {\flushright}                                                          

 \title{Magnetic Monopole Decay and Its Consequences}
\author{Sherman Frankel}
\address{\it Department of Physics and Astronomy}
\address{ \it University of Pennsylvania}
\date{May 28, 1999}

\maketitle

\begin{abstract}
         We return to further examination of the need for the
existence of magnetic monopoles based on the decades old derivation
  [Sherman Frankel, Amer. Jour. of Physics, {\bf 44} 7 683-6 1976],
of both  the
monopole-charge force, $\vec F = eg(\vec r\times\vec v/c)/r^3$, as well as the Dirac
angular momentum
, $\vec l_f = eg \hat r$, from $P$ and $T$ conservation alone,
 without
 any recourse whatsoever to Maxwell's Equations.
 The $eg$ product also appeared in
the charge conserving equation, $deg/dt = 0 = edg/dt + gde/dt$,
and this paper examines the
 second possible
solution, $dg/dt = - de/dt$, namely monopole decay, accounting for the
non-observance of stable monopoles.
  It treats  astrophysics
and neutrino physics consequences and suggestions for experimental
searches.
\end{abstract}

\newpage

\vsl
\centerline {\bf           I. Introduction    }

\vss
   Over two decades have passed since I published the paper entitled
 `` Pseudoscalar and Scalar Charges'' which is the basis
for this present work \cite{www}.
In that paper I
showed that imposition of parity conservation and time reversal invariance
allowed one to {\it derive} the fundamental relations dealing with monopoles
(charge = g)  and electric charges (charge = e) {\it without the slightest
knowledge of any of the laws of electromagnetism.} This was accomplished 
by
requiring that 
charges are scalars, even under $T$, and monopole charges are pseudoscalars,
odd under T \cite{agreement}.

The proof  was carried out for spinless particles by writing down the
most general force and angular momentum in terms of the relative
velocities, $\vec v$, and separations, $\vec r$, the charges
 $e$ and $g$, and demanding
$P$, $T$, angular momentum, linear momentum, and charge conservation.
As we showed in reference~\cite{www},  the expressions for the most general force and
most general angular momentum each contain a number of possible terms
consistent with the symmetry assumptions for $e$ and $g$.

The only possible term that did not vanish
               was the well known force \cite{finitephoton},

falling off {\it exactly} as
$1/r^2$, namely:

\begin{equation}
 \vec F = e g (\vec r \times \vec v/c)/r^3 
\label{equation1}
\end{equation}

However, what was most revealing was that, in addition to the orbital
angular momentum $\vec r \times \vec p$, one automatically found that a
 new term, which can be called the ``field angular momentum'', needed to
 be included to conserve overall angular momentum. (This came about because the
charge-monopole force of Equation~\ref{equation1} is not a central force and therefore the
orbital angular momentum alone could not be a constant of the motion.)
It was found that the surviving term that was needed for angular momentum
 conservation
was just
\begin{equation}
 \vec l_f = eg      \hat r 
\label{equation2}
\end{equation}
where $\hat r $ was the {\it unit vector} separating the charges.

This is the expression
\cite{Dirac}
which Dirac used as the basis for his argument that existence of a single
monopole would result in a $\it quantized$ electric charge. A classical 
derivation
was given by H. A. Wilson \cite{Wilson}
Yet our
derivation of Equation~\ref{equation2}
made no use of  Equation~{\ref{equation1} or any part of electromagnetic
theory.

         The quantization of the angular momentum by the expression
\begin{equation}
\mid \vec l_f \mid  = eg  = n \hbar
\label{equation3}
\end{equation}
with n = a ratio of integers,
proved  that the {\it product} of the charges was quantized, not that
they were separately quantized \cite{quantized}.

Equations~\ref{equation1} and ~\ref{equation2}
 depend on the {\it product} of the charges and it is useful to note that
their derivation also required that the
{\it product of the
charges be conserved }, namely:
\begin{equation}
 d(ge)/dt = 0 = gde/dt + edg/dt 
\label{equation4}
\end{equation}

As described more fully in ref. 1,
this came about naturally
since it was necessary to include every possible term
having  the dimensions and $P$ and $T$ properties in a total angular momentum and
require  that the torque be given by $\vec T = d\vec L_{total}/dt $.
Further, the derivation had to include charge conservation and not
add it in as an ad hoc assumption in order to obtain  eqs. 1) and 2).

         There are two possible solutions to this generalized equation:

 a)
$de/dt$ and $dg/dt$ {\it separately} vanish, which is the usual conventional 
assumption
that yields stable charges.

 b)
the {\it sum} vanishes and one charge, presumably the monopole, $g$, can decay
into an ordinary charge \cite{timedecay}

         This present paper returns to examine the consequences of this
``generalized charge
conservation'', bringing to bear considerations of the neutrality and
isotropy of the universe and, also, the
consideration of whether there exists a preferred
direction in space. It proposes a set of experimental consequences
relating to the existence of a possible fourth neutrino, the possibility
of a new neutral current contribution to the weak breakup of deuterium
in the SNO experiment, and the possibility of the observance of large
photon jets in the universe, in one scenario accompanied by neutrinos.
  It also examines unique
ways to search for monopole production at accelerators.

\vss
\centerline { \bf II.  Magnetic Monopole Decay  }

\vss
         The first point I shall address is the astrophysical constraint
on the ``classical" derivation of charge {\it quantization}. Two approaches, one
based on
the angular momentum in the electromagnetic field, $\int \vec r \times
(\vec E \times \vec B) dV $ \cite{Wilson}
 and the other, based on the angular momentum
acquired in the impulse approximation for the scattering of a charge by
a fixed monopole \cite{Goldhaber}
 are well known to lead to Equation~\ref{equation2}.
with its unique unit vector relationship.

          However the classical charge quantization argument
assumes that a monopole and charge exist but that the rest of the universe
is isotropic and uncharged. For example, if there were only one other
unpaired electron somewhere in space, the angular momentum would now
be $ \vec l_f = ge_1 \hat r_1 + ge_2 \hat r_2$ ,which is not a unit vector.
Thus the
semi-quantum mechanical derivation of charge quantization would not be
independent of
particle position and would fail.
However it is possible to restore the charge quantization by assuming
an isotropic and uncharged universe and {\it then} creating an
 electric-monopole pair from the vacuum, which is  allowed by 
generalized charge
conservation. This would result in the existence of a single $eg$ pair plus an
isotropic
and uncharged remaining universe and the semi-quantum mechanical derivation
could then proceed.
Full  quantum-mechanical treatments of the $eg$ quantization have been given
by several authors \cite{WuYang}.

\vsp

\vss
\centerline {\bf III. Monopole Production }

\vsp
 It is crucial to recognize that both the production and decay of
magnetically charged particles
will be accompanied by jets of photons. Unlike ordinary radiative
interactions, (both production and decay),
determined by the electric fine structure constant, $e^2/\hbar c$,
such production or decay
       of magnetically charged particles would be dominated by the
huge monopole fine structure constant, $g^2/\hbar c$,
 which would be at least $(137)^2$
times larger than $e^2/\bar{h} c$.  These would make interesting and dramatic
signatures independent of whether monopoles decayed.
 It will be convenient to think that
a monopole charge particle would recoil against the momentum of the radiated
 photons so that
the photon distribution will not be isotropic but directed in a jet-like cone.

In order not to violate any known conservation laws, the most likely
reactions would involve simultaneously producing a magnetically charged
particle and its antiparticle, e.g.,

\begin{equation}
p + \bar{p} \rightarrow    M + {\bar{M}}  + {\rm photon \ jet}
\label{equation5}
\end{equation}

\begin{equation}
e + \bar{e}  \rightarrow  M + {\bar{M}} + {\rm photon \  jet}
\label{equation6}
\end{equation}

\begin{equation}
\gamma + {\rm nucleon} \rightarrow  M + {\bar{M}}  + {\rm nucleon}  + {\rm
photon \ 
jet}
\label{equation7}
\end{equation}

Here $M$ designates either a magnetically charged baryoM or leptoM.

\vsp
a) First,   we consider a most interesting possibility, the leptoM, $l^M$,
undergoing leptonic decay.

The production mechanism could be that of Equation~\ref{equation5} with $M$ being a leptoM.

A most interesting decay possibility would then be the conventional weak decay:

\begin{equation}
l^M \rightarrow l + \nu_l + \bar{ \nu_M} + {\rm photon \  jet}
\label{equation8}
\end{equation}

where $l$ denotes electron, muon, or tau and $\nu_M$ is the
 monopole
neutrino.

\vsp  a) We next  consider monopole charged baryons,
 (baryoMs)

  If the the baryoMs  were to have the
same baryon number as the proton, they could decay by either strong or
weak interactions. The strong interaction would dominate so the decay
reaction
could be:
\begin{equation}
M   \rightarrow p + x + {\rm photon \  jet}
\label{equation9}
\end{equation}

Each $ M$ or $\bar M$  would decay, producing protons or antiprotons
 plus their  jets of photons.

The decays of the baryoMs would  then appear to be oppositely directed hadrons and
photon jets. Thus the final state would have four jets plus  a proton and
antiproton and probably some additional bosons.

\vsp
Another possibility is the production of single
monopole charged bosons
(bosoMs) using a high energy pion beam, in a fixed target accelerator.
 The production process would not conserve ordinary charge,
following the inverse transformation, $e \rightarrow  g $, for example in
 the reaction
\begin{equation}
 \pi^+  + p \rightarrow bosoM   + p + x +  {\rm photon \  jet}
\label{equation}
\end{equation}

That  bosoM  ( perhaps a $\pi^M$) could then decay back by the reaction:
\begin{equation}
 \pi^M  \rightarrow  \pi^-  + x  + {\rm photon \ jet}
\label{equation11}
\end{equation}

where x could be  neutral boson configurations.

\vsp

\vss
\centerline{ \bf IV. Astrophysical Consequences   }

\vsp
\centerline {\bf A. The Missing Mass of the Universe }

\vsp
 If very
heavy monopole pairs were created in the early universe, there
would now be a background of monopole neutrinos and antineutrinos arising
from the leptonic
monopole decay. From studies of the Z vector boson decays, such a fourth
neutrino would have to have a mass greater than 45 Gev, but only
 if the leptom
     were
coupled to the Z in the same way as ordinary leptons.  But if this were not the
coupling, there would be no constraint on the monopole neutrino mass.

 Since we have no
{\it a priori} knowledge of the mass of the monopole we cannot estimate the
number of such neutrinos in the present universe, nor can we estimate
the contribution of this source of mass to the ``missing mass'' of the
universe if the monopole neutrino were not massless. If one assumed that
the monopole leptonic decay was governed by the strength of the usual weak
coupling constant and one knew the missing mass, one could calculate what
pairing of monopole mass and monopole neutrino mass could account for the
missing mass.

\vss 
\centerline  {\bf B. Neutral Monopole Neutrino Currents }

\vsp
     Monopole decays                 would produce monopole
neutrinos
which could interact in   neutrino detectors by the reaction
$\nu_g + nucleon \rightarrow \nu_g + nucleon$, raising the neutral current
rate so that neutrino disintegration of deuterium would be enhanced. Thus
the reaction:

\vsp
6)     $\nu_m + d \rightarrow n + p + \nu_m $ might be observed in the
SNO experiment and be mistaken for an ordinary neutral current anomaly.
Such events would also come from the ordinary neutrinos from
the monopole decays.

\vss   
\centerline{\bf C. Photon Bursts }

\vsp  If somewhere in the universe monopoles and antimonopoles are
being created and decay there will be regions from which large,
high energy, photon bursts would be observed. If these monopoles were leptoMs
they would be accompanied by
neutrinos, both standard neutrinos and monopole neutrinos.
This correlation, if observed, would suggest the production of $g$ and
$\bar{g}$ lepton pairs. Thus one would look in neutrino detectors for
coincidences with photon bursts.

\centerline {\bf  D. A Preferred Direction in Space?}

      Imagine an electron and positron approaching each other in their cm system and
annihilating to produce a monopole boson and a charged boson in accordance with
generalized charge conservation and producing the field angular momentum $ge$ times
their unit vector separation.
 It is of course possible that the universe is not exactly
spatially isotropic. If it came into being from a state of zero angular
momentum,
the universe at that instant would have a field angular momentum
$\sum g_i e_j \vec r_{ij}/r_{ij}$.       To conserve overall angular momentum,
the universe
would then have to possess an orbital angular momentum to compensate for the
field  angular momentum. Thus one might wish to attempt to refine measurements
searching for an orbital angular momentum of the universe.

\newpage
 \centerline{\bf V. Accelerator Experiments  }

\vsp
We cannot reliably predict the monopole masses. But,
      if monopoles were not too massive, they could be made at Fermilab
by the allowed reaction $q  + \bar{q} \rightarrow M + \bar{M} $. Because
of the huge monopole coupling constant, one would expect photon jets at
opposite $\phi$ as the $M$ and $\bar{M}$ recede. In addition, one would
get photon jets from each of the monopole decays. Unfortunately, in the
case of monopole leptoMs, there would be four undetected neutrinos and
a large missing mass in the event, making reconstruction difficult.
If the jet energies were very large, the monopoles might have very low
energies before their decay, somewhat simplifying the analysis.

However, the production of a single bosoM would be a  simpler
event to look for and reconstruct.
  The reaction might be $p  + \bar{p}  \rightarrow $
bosoM + photon jet + singly charged pion, the jet possibly radiated in
the bosoM direction. The bosoM could then decay back by: bosoM $\rightarrow
\pi $ + photon jet. Thus, in this case,  one would expect to see
a jet roughly in the bosoM direction opposite to a pion and then
another jet opposite the bosoM decay pion. In addition the pions would have
{\it opposite } sign charges.
 In this simple case there
would be no missing energy and a unique spatial configuration of two
pion-jet pairs.

\vss
\centerline {\bf  VI. Discussion }

\vsp
        Derivation of the fundamental laws of electrical interactions from
the $P$, $T$  invariance properties of electric and magnetic charges alone has
allowed
for the possibility for monopole decay. This would account for the
non-observance of stable magnetic monopoles in over a half-century of 
energetic search.
  Including such generalized
charge conservation and monopole decay in a full theory
cannot be accomplished by merely inserting monopoles
into the classical Maxwell Equations which  were constructed to account
for a world containing electric charges and no experimental evidence for monopoles.
It will require a quantum-mechanical theory, perhaps an
electro-weak theory  including new $W$'s and $Z$'s, but we simply cannot predict its
structure.
Nor can electromagnetic theory, based on observed
electromagnetic effects in a monopole-free world, be trivially modified to incorporate
 monopole decay.
   In this paper we have examined the decays of possible monopole charged particles,
leptons and baryons, without discussing how they are related, the size of the 
families, the possible quark constituents,  etc. This is an interesting area for
speculation {\it even if} monopoles existed and monopole decay did not exist.
 Further, we are well aware that this is a classical theory
and that it is not obvious how one will have to
    modify present theories to include monopole decay just as one cannot now make new
theories that include monopoles without accounting for their absence as stable particles.
Our thought is that this paper might stimulate such efforts.
Finally, we
propose {\it experimental} searches for effects of monopole photon bursts, monopole
neutrinos, monopole production at Fermilab, and a possible assymetry in space that could
be carried out with existing detectors..

\vss Acknowledgements:
We wish to thank many theoretical and experimental colleagues for 
their incisive comments on this work.

\fr  monolate5 printed \today

\begin{thebibliography}{99}
\bibitem{agreement} The choice of name is arbitrary but in agreement
with Maxwell's Equation: div $\vec E = \rho_e $.
\bibitem{quantized} Of course if one assumed that only a
single monopole existed in the universe, it would follow that electric
charges were quantized, but then one would be faced with wondering why 
nature bothered to produce only a single monopole.
\end{thebibliography}
\end{document}